\documentclass[aps,twocolumn,secnumarabic,balancelastpage,amsmath,amssymb,nofootinbib,hyperref=pdftex,prb]{revtex4-1}

\usepackage{graphics}
\usepackage[pdftex]{graphicx}
\graphicspath{{Figures/}}
\usepackage{longtable}
\usepackage{amsmath}
\usepackage{hyperref}

\makeatletter
\def\p@subsection{\thesection\,}
\makeatother

\newcommand{\bo}[1]{{\bf #1}}
\newcommand{\avg}[1]{\left\langle #1 \right\rangle}
\newcommand{\pone}{\hat{\tau}_1}
\newcommand{\ptwo}{\hat{\tau}_2}
\newcommand{\pthree}{\hat{\tau}_3}

\begin{document}

\title{Zero-energy bound state at the interface between an $s$-wave superconductor and a disordered normal metal with repulsive electron-electron interactions}
\author{Christopher R. Reeg}
\author{Dmitrii L. Maslov}
\affiliation{Department of Physics, University of Florida, 
 P. O. Box 118440,
Gainesville, FL 32611-8440, USA}
\date{\today}
\begin{abstract}
In recent years, there has been a renewed interest in the proximity effect due to its role in the realization of topological superconductivity. Here, we study a superconductor--normal metal proximity system with repulsive electron-electron interactions in the normal layer. It is known that in the absence of disorder or normal reflection at the superconductor--normal metal interface, a zero-energy bound state forms and is localized to the interface [Fauch\`{e}re \emph{et al.}, Phys. Rev. Lett. {\bf 82}, 3336 (1999)]. Using the quasiclassical theory of superconductivity, we investigate the low-energy behavior of the density of states in the presence of finite disorder and an interfacial barrier. We find that as the mean free path is decreased, the zero-energy peak in the density of states is broadened and reduced. In the quasiballistic limit, the bound state eliminates the minigap pertinent to a noninteracting normal layer and a distinct peak is observed. When the mean free path becomes comparable to the normal layer width, the zero-energy peak is strongly suppressed and the minigap begins to develop. In the diffusive limit, the minigap is fully restored and all signatures of the bound state are eliminated. We find that an interfacial potential barrier does not change the functional form of the density of states peak but does shift this peak away from zero energy.
\end{abstract}

\maketitle

\section{Introduction} \label{Intro}

Some time ago, Fauch\`{e}re \emph{et al.} \cite{Fauchere:1999} showed that in the highly idealized limit of no disorder and perfect transmission, a zero-energy bound state arises at the interface between an $s$-wave superconductor and a normal metal with repulsive electron-electron interactions. The one-dimensional analog of this effect is the zero-energy peak in the density of states of a Luttinger-liquid quantum wire in proximity to a superconductor \cite{Winkelholz:1996} and, even in the absence of a superconductor, of a Luttinger liquid with spatially modulated strength of the repulsive interaction.\cite{safi:1995, *safi:1995b,   *safi:1996, maslov:1998} The (higher-dimensional) zero-energy state was originally invoked as an explanation of reentrant paramagnetism observed by Visani \emph{et al.} in normal-metal coated superconducting cylinders; \cite{Visani:1990} however, alternative explanations of this effect within the single-particle picture have also been suggested (see Ref.~\onlinecite{gogadze:2010} for a review). To the best of our knowledge, there has yet to be a direct experimental observation of the zero-energy peak at a superconductor-normal metal interface; this peak is distinct from the well-known phenomenon of reflectionless tunneling, which produces a zero-bias peak in the conductance \cite{Kastalsky:1991,vanWees:1992,Magnee:1994} but not in the density of states. \cite{Volkov:1994} It has even been suggested that such a peak is an artifact of the quasiclassical approximation and does not occur at all in the fully quantum-mechanical treatment of the problem. \cite{Valls:2010} Relatedly, the most convincing experiments to date involving the search for Majorana fermions have shown a zero-energy peak in the density of states at the interface between a topological superconductor and normal metal; \cite{Mourik:2012,Deng:2012,Das:2012,Finck:2013} these developments thus necessitate a better understanding of the non-Majorana zero-energy state as well.

In this paper, we study the zero-energy bound state in a more realistic context. In particular, we study how disorder and finite normal reflection (in addition to Andreev reflection) at the superconductor--normal metal interface affect the low-energy behavior of the local density of states. We show that impurity scattering eventually transforms a zero-energy singularity in the density of states in the ballistic limit into a hard mini-gap in the diffusive limit. The zero-energy peak is quite sensitive to disorder; our solution of the Eilenberger equations shows that the peak is severely suppressed compared to the ballistic case even if the mean free path ($\ell$) is five times longer than the normal metal thickness ($d$). When $\ell=d$, the peak is barely discernible. This may explain why the peak has not been observed by scanning tunneling microscopy on junctions between superconductors and conventional metals, as thin metallic films are inevitably disordered. On the other hand, normal reflection at the interface shifts the zero-energy peak in the density of states to a finite energy but does not smear the peak. Nevertheless, the amplitude of the peak is reduced, as normal reflection suppresses the pairing amplitude induced in the normal metal. We also address the claim of Ref.~\onlinecite{Valls:2010} that no zero-energy bound state arises in this system if one solves the Bogoliubov--de Gennes (BdG) equation directly, without invoking the quasiclassical approximation. We do find the zero-energy peak by solving the BdG equation.

\begin{figure}[t!]
\includegraphics[width=\linewidth]{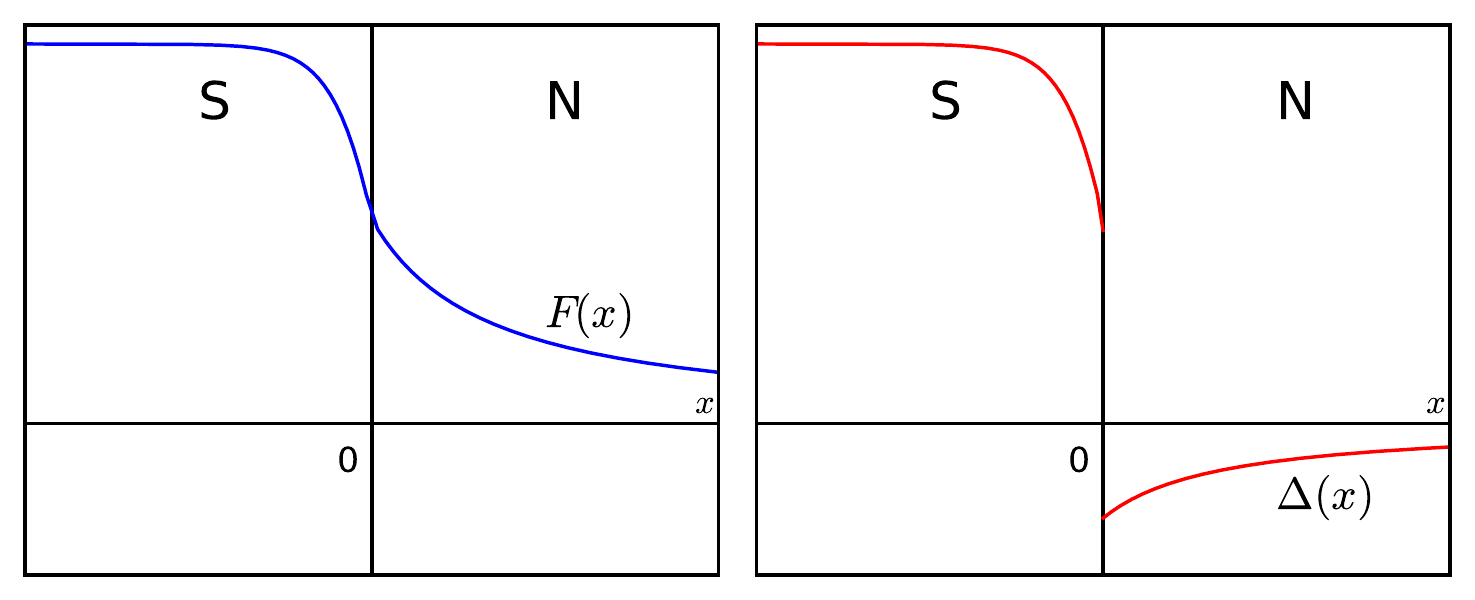}
\caption{\label{sketch} (Color online) The pairing amplitude $F(x)$ (left curve) and pairing potential $\Delta(x)$ (right curve) profiles near the (ideal) interface between a superconductor and a normal metal with repulsive electron-electron interactions ($\lambda_N>0$).}
\end{figure}

A normal metal (N) placed in good contact with a superconductor (S) inherits superconducting correlations expressed through a nonvanishing pairing amplitude, $F(x)=\avg{\psi_\uparrow(x)\psi_\downarrow(x)}$, where $x$ is the coordinate perpendicular to the SN interface. However, electron-electron interactions are required for the normal metal to inherit a non-zero pairing potential, $\Delta(x)=-\lambda(x)F(x)$, with $\lambda$ the electron-electron interaction coupling. The coupling $\lambda$ can take positive values (repulsive interaction) or negative values (attractive interaction) in the normal metal, depending on the balance between Coulomb repulsion and phonon-mediated attraction. It was first noted by de Gennes that repulsive interactions in the normal metal induce a sign change in $\Delta(x)$ at the SN interface. \cite{deGennes:1964} Generic spatial profiles of the pairing amplitude and pairing potential are sketched in Fig.~\ref{sketch}.

Any infinite system with a pairing potential that has opposite signs at $x=\pm\infty$ will harbor a zero-energy bound state. \cite{Kosztin:1998, Adagideli:1999} To see this in the context of an SN proximity system, we start with the BdG equation, \cite{deGennes}
\begin{equation} \label{BdG}
\bigl[H_0\pthree+\Delta(x)\pone\bigr]\hat{\psi}_E({k}_\perp,x)=E\hat{\psi}_E({k}_\perp,x),
\end{equation}
where $H_0=-\partial_x^2/2m+k_\perp^2/2m-\mu$, $\hat{\psi}_E({k}_\perp,x)=[u_E(k_\perp,x),v_E(k_\perp,x)]^T$ is a particle-hole spinor wave function with energy $E$, and $\hat{\tau}_i$ are Pauli matrices in Nambu space. The pairing potential $\Delta(x)$ is taken to be a real function. In the quasiclassical approximation, the wave function is represented as a product of a rapidly oscillating factor and a slowly varying envelope function: $\hat{\psi}_E({k}_\perp,x)\to\hat{\phi}_E({k}_{F,\perp},x)\exp(ik_{F,x}x)$, where $k_{F,\perp}^2+k_{F,x}^2=k_F^2$. If the second derivative of $\hat{\phi}$ is neglected, Eq.~(\ref{BdG}) simplifies to the Andreev equation, \cite{Andreev:1964}
\begin{equation} \label{Andreev}
\bigl[-iv_x\partial_x\pthree+\Delta(x)\pone\bigr]\hat{\phi}_E(v_x,x)=E\hat{\phi}_E(v_x,x),
\end{equation}
where $v_x=\bo{v}_F\cdot\hat{x}$. The special property of Eq.~(\ref{Andreev}) is that it admits a bound state solution at $E=0$ if the pairing potential changes sign at the SN interface. For definiteness, let us consider a semi-infinite superconductor ($x<0$) in contact with a semi-infinite normal metal ($x>0$). Then the solution to Eq.~(\ref{Andreev}) at $E=0$ that is bounded in both the superconductor and normal metal is given by
\begin{equation} \label{BdGsolution}
\hat{\phi}_0(v_x,x)=C\left(\begin{array}{c} 1 \\ i \end{array}\right)\exp\left[\frac{1}{v_x}\int_0^x\Delta(x')dx'\right].
\end{equation}
Here, $v_x$ is taken to be positive and $C$ is a normalization constant. For $x\to-\infty$ (i.e., deep into S), the wave function decays exponentially. Because the pairing potential decays into the normal metal as $\Delta(x)\propto-\lambda_N/x$ far from the SN interface at zero temperature \cite{Alexandrov:2008} (this will be discussed in more detail in Sec.~\ref{Quasiclassics}), the envelope of the wave function decays as a power law, $\hat{\phi}_0(v_x,x)\propto1/x^\beta$ for $x\to+\infty$ (i.e., deep into into  $N$), with exponent $\beta\propto\lambda_N$. Thus, based on the quasiclassical argument, we expect there to be a zero-energy bound state localized to the interface between a superconductor and a normal metal with repulsive interactions. Note that the very existence of the bound state does not depend on any details of the functional form of $\Delta(x)$ other than that it must change sign somewhere between S and N. This point will be important in Sec.~\ref{Beyond QC}, where we go beyond the quasiclassical approximation.

The remainder of the paper is organized as follows. In Sec.~\ref{Quasiclassics}, we review the quasiclassical formulation of superconductivity as it pertains to the problem at hand. We then solve the quasiclassical equations analytically using a simple step-function model for $\Delta(x)$ and calculate the local density of states in Sec.~\ref{Step}. In particular, we solve this model in the ballistic limit assuming perfect transmission at the SN interface in Sec.~\ref{Ballistic}, we consider the effects of a tunnel barrier in the ballistic limit in Sec.~\ref{Barrier}, and we discuss the diffusive limit in Sec.~\ref{Diffusive}. In Sec.~\ref{Numerical}, impurity scattering is taken into account and a fully self-consistent calculation of both the pairing potential and the local density of states is presented for various values of mean free path. Our numerical methods are discussed in Sec.~\ref{Methods}, while the results of the calculation are given in Sec.~\ref{Results}. Finite temperatures are discussed in Sec.~\ref{FiniteT}. In Sec.~\ref{Beyond QC}, we show numerically that the zero-energy bound state can be obtained from nonquasiclassical methods as well. Our conclusions are given in Sec.~\ref{Conclusions}.

\begin{figure}[t!]
\includegraphics[width=\linewidth]{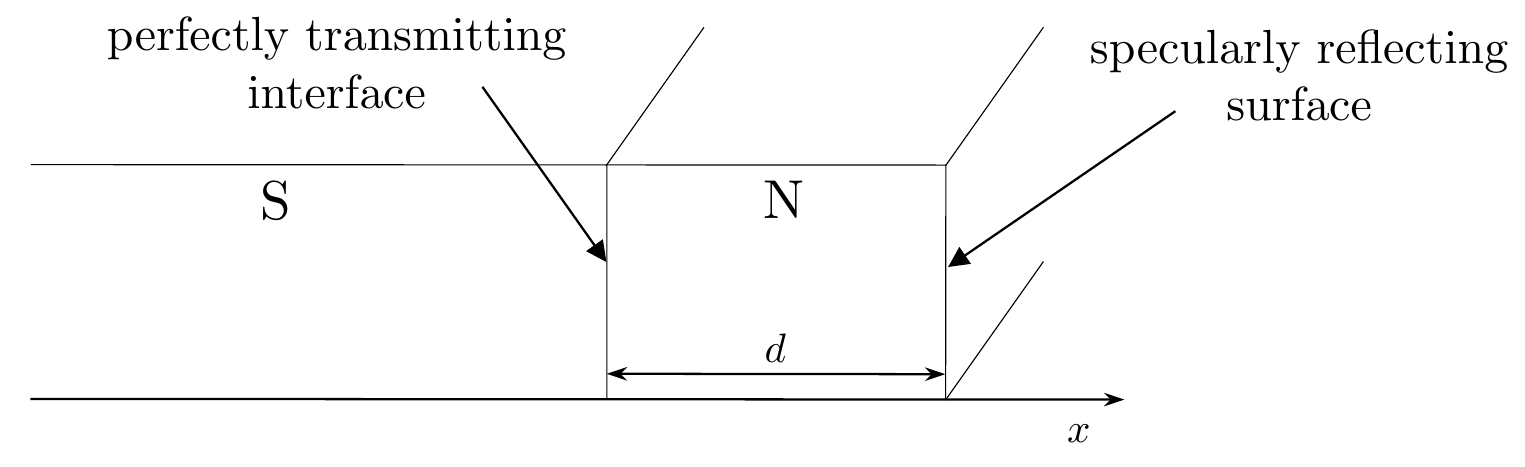}
\caption{\label{setup} Cross section of SN proximity geometry. Interface located at $x=0$ is taken to be perfectly transmitting throughout most of the paper, while vacuum-normal metal boundary at $x=d$ is assumed to be specularly reflecting. System is infinite in both the $y$ and $z$ directions.}
\end{figure}

\section{Quasiclassical Theory} \label{Quasiclassics}
Throughout the remainder of this paper, we consider a semi-infinite superconductor, located at $x<0$, in contact with a normal metal of thickness $d$, located at $0<x<d$ (Fig.~\ref{setup}). Both materials are infinite in the directions transverse to the SN interface. We assume that the vacuum-normal metal boundary is specularly reflecting, and throughout most of the paper we take the SN interface to be perfectly transmitting. However, in Sec.~\ref{Barrier} we do allow for interfacial scattering.

Assuming (as in Sec.~\ref{Intro}) that the Green's functions vary slowly on the Fermi wavelength scale and integrating out the momentum dependence of the Gor'kov Green's functions (which has the effect of projecting $k\to k_F$) allows one to rewrite the Gor'kov equations \cite{AGD} in a  greatly simplified form. These simplified equations are the Eilenberger equations,\cite{Eilenberger:1968} which can be expressed compactly as a single matrix equation,
\begin{equation} \label{Eilenberger}
-v_x\partial_x\hat{g}(v_x,\omega,x)=\bigl[\omega\hat{\tau}_3+\Delta(x)\hat{\tau}_1+\hat{\sigma}(\omega,x),\hat{g}(v_x,\omega,x)\bigr].
\end{equation}
In Eq.~(\ref{Eilenberger}), $\hat{g}$ is a $2\times2$ quasiclassical matrix Green's function containing both normal (diagonal) and anomalous (off-diagonal) components, $\hat{g}=g\pthree+f(\pone+i\ptwo)+f^\dagger(\pone-i\ptwo)$, $[\hat A,\hat B]=\hat A\hat B-\hat B\hat A$, and $\hat{\sigma}(\omega,x)$ is a matrix self-energy due to impurity scattering. The Green's functions obey the normalization condition $\hat{g}^2(v_x,\omega,x)=\hat{1}$.

The self-consistency condition on the pairing potential is given by
\begin{equation} \label{delta}
\Delta(x)=-\pi\lambda(x) N_0\int\frac{d\omega}{2\pi}\avg{f(v_x,\omega,x)},
\end{equation}
where $N_0=mk_F/2\pi^2$ and $\avg{\cdots}=\int_{-1}^1 d\zeta/2$ denotes an angular average over the Fermi surface. Here, we introduce the shorthand notation $\zeta=v_x/v_F$. Using the anomalous Green's function $f(v_x,\omega,x)$ in the non-interacting case (i.e., $\lambda_N=0$), it is straightforward to show from Eq.~(\ref{delta}) that the pairing potential decays as $\Delta(x)\sim-\lambda_N N_0v_F/x$ for $x\gg\xi_S$, where $\xi_S$ is the coherence length of the superconductor.

Impurity scattering gives rise to a self-energy $\hat{\sigma}(\omega,x)$ that must be calculated self-consistently. In the Born approximation, the self-energy is expressed as
\begin{equation} \label{selfenergy}
\hat{\sigma}(\omega,x)=\frac{v_F}{2\ell}\avg{\hat{g}(v_x,\omega,x)},
\end{equation}
where $\ell$ is the mean free path. The local density of states (LDOS) is given by
\begin{equation} \label{LDOS}
N(E,x)=N_0\text{Re}\left[\avg{g^R(v_x,E,x)}\right],
\end{equation}
where the retarded Green's function $g^R(v_x,E,x)$ is found by continuing $\omega\to-iE+\delta$. Further details about the self-consistent calculation of the self-energy are discussed in Sec.~\ref{Numerical}.

\section{Step Model for $\Delta(x)$} \label{Step}
\begin{figure}[t!]
\includegraphics[width=0.6\linewidth]{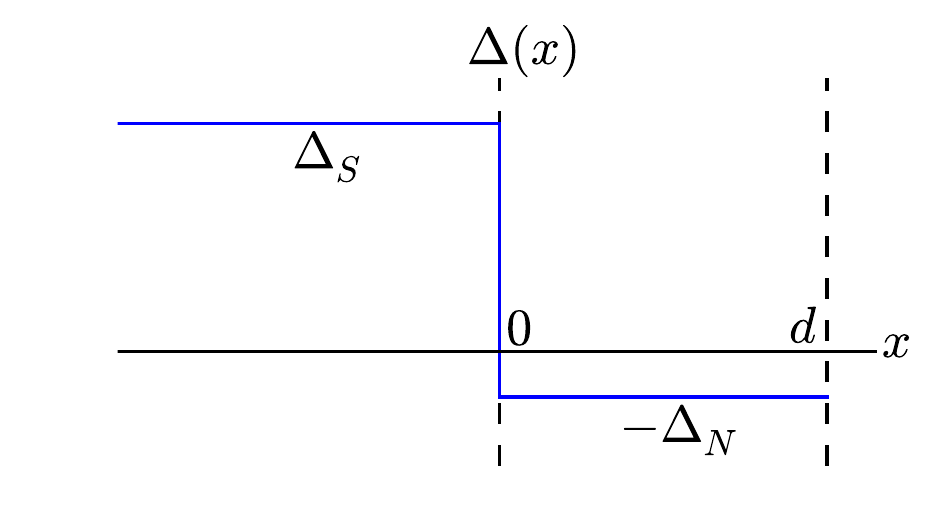}
\caption{\label{stepmodel} (Color online) Step model for $\Delta(x)$.}
\end{figure}

We first show analytically how the expected zero-energy state occurs in this system. To this end, we approximate the pairing potential by a step function (see Fig.~\ref{stepmodel}),
\begin{equation}
\Delta(x)=\left\{\begin{array}{lcc}\Delta_S, & & x<0, \\ \\
	-\Delta_N, & & 0<x<d.
	\end{array}\right. 
\end{equation}
This approximation allows us to solve Eq.~(\ref{Eilenberger}) in both the ballistic and diffusive limits. This model was analyzed previously in the ballistic limit by Fauch\`{e}re \emph{et al.}; \cite{Fauchere:1999} for the sake of completeness, we will review some of those results as well as add the effects of a potential barrier at the SN interface. In particular, we obtain an analytic form of the density of states at all energies and at any point within N, and extract an explicit form of the peak near $E=0$. We will also study the diffusive limit within this model.

\subsection{Ballistic Limit 
without 
Interfacial Barrier} \label{Ballistic}
We first consider the ballistic limit ($\ell\to\infty$) and assume perfect transmission of the SN interface, in which case Eq.~(\ref{Eilenberger}) is readily solved in both the S and N regions:
\begin{equation} \label{Eilenberger_sol}
\begin{aligned}
\hat{g}_N(\pm v_x,\omega,x)&=\frac{1}{{\Omega}_N}\biggl[A_1(\omega\hat{\tau}_3-\Delta_N\hat{\tau}_1)+ \\
	&+A_2(\Delta_N\hat{\tau}_3+\omega\hat{\tau}_1\mp i{\Omega}_N\hat{\tau}_2)e^{2{\Omega}_Nx/|v_x|}+ \\
	&+A_3(\Delta_N\hat{\tau}_3+\omega\hat{\tau}_1\pm i{\Omega}_N\hat{\tau}_2)e^{-2{\Omega}_Nx/|v_x|}\biggr], \\
\hat{g}_S(\pm v_x,\omega,x)&=\frac{1}{{\Omega}_S}\biggl[(\omega\hat{\tau}_3+\Delta_S\hat{\tau}_1)+ \\
	&+B(\Delta_S\hat{\tau}_3-\omega\hat{\tau}_1\pm i{\Omega}_S\hat{\tau}_2)e^{2{\Omega}_Sx/|v_x|}\biggr],
\end{aligned}
\end{equation}
where ${\Omega}_{N(S)}^2=\Delta_{N(S)}^2+\omega^2$. This form of the solution is chosen so as to explicitly satisfy the symmetries of the Eilenberger equation, $f(v_x,\omega,x)=f^\dagger(-v_x,\omega,x)$ and $g(v_x,\omega,x)=g(-v_x,\omega,x)$. The coefficients are determined by enforcing suitable boundary conditions. Specular reflection at the vacuum--N boundary requires $\hat{g}_N(v_x,\omega,d)=\hat{g}_N(-v_x,\omega,d)$, and perfect transmission at the SN interface implies $\hat{g}_S(v_x,\omega,0)=\hat{g}_N(v_x,\omega,0)$. In the limit of both $\omega$ and $\Delta_N$ being much smaller than $\Delta_S$, the normal Green's function can be expressed as
\begin{equation} \label{Green's function}
g_N(v_x,\omega,x)=\frac{\omega}{{\Omega}_N}\frac{{\Omega}_N\sinh{\chi}-\Delta_N\cosh{\chi}+\Delta_N\cosh(\chi \tilde{x})}{\Omega_N\cosh{\chi}-\Delta_N\sinh{\chi}},
\end{equation}
where we have defined 
\begin{equation}
{\chi}=2{\Omega}_N d/|v_x|
\label{chi}
\end{equation}
and
\begin{equation}
\tilde{x}=1-x/d.
\label{tildex}
\end{equation}

After analytic continuation, the retarded Green's function has a non-zero real part due to its poles. Focusing only on energies $E>0$,
\vfill
\begin{widetext}
\begin{equation} \label{realpart}
\begin{aligned}
\text{Re}\,g_N^R(v_x,E,x)&=\pi\left(\sinh\chi-\frac{\Delta_N}{\Omega_N}\cosh\chi+\frac{\Delta_N}{\Omega_N}\cosh(\chi\tilde{x})\right)\delta(\chi-\bar{\chi})\theta(\Delta_N-E) \\
	&+\pi\sum_n\left(\sin\chi+\frac{\Delta_N}{\Omega_N}\cos\chi-\frac{\Delta_N}{\Omega_N}\cos(\chi\tilde{x})\right)(-1)^n\delta(\chi-\chi_n)\theta(E-\Delta_N),
\end{aligned}
\end{equation}
\end{widetext}
where we have defined
\begin{subequations} \label{chi2}
\begin{eqnarray}
\bar{\chi}&=&\tanh^{-1}(\Omega_N/\Delta_N);\\
\chi_n&=&\tan^{-1}(\Omega_N/\Delta_N)+n\pi.
\end{eqnarray}
\end{subequations}
(It is understood that now $\Omega_N^2=\Delta_N^2-E^2$ for $E<\Delta_N$ and $\Omega_N^2=E^2-\Delta_N^2$ for $E>\Delta_N$.) The sum in Eq.~(\ref{realpart}) runs over all $n>(2\Omega_Nd/v_F-\chi_0)/\pi$.

To determine the form of the LDOS near zero energy, we expand in the limit $E\ll\Delta_N$.  
Changing  the angular integration variable from $\zeta$ to $\chi$, we rewrite the LDOS as
\begin{equation} \label{FirstExpansion}
\begin{aligned}
N(E,x)&=\frac{2\pi N_0\Delta_Nd}{v_F}\int_{\frac{2\Delta_Nd}{v_F}}^\infty\frac{d\chi}{\chi^2}\biggl(-e^{-\chi} + \\
	&+ \cosh(\chi\tilde{x})\biggr)\delta\bigl(\chi-\ln[2\Delta_N/E]\bigr).
\end{aligned}
\end{equation}
If $(\Delta_N/E)^{\tilde{x}}\gg1$, we can neglect $\exp(-\chi)$ compared to $\exp(\chi)$. Expanding further in this limit, the integral in Eq.~(\ref{FirstExpansion}) evaluates to
\begin{equation} \label{AnalyticLDOSlowenergy}
N\left(E,x\right)=\frac{\pi N_0\Delta_Nd/v_F}{\ln^2(2\Delta_N/E)}\left(\frac{2\Delta_N}{E}\right)^{\tilde{x}}.
\end{equation}
Equation (\ref{AnalyticLDOSlowenergy}) describes a singularity in the LDOS at zero energy due to the presence of the bound state, the amplitude of which is determined by the parameter $\Delta_Nd/v_F$. While this singularity is present everywhere inside the normal metal, it weakens away from the SN interface. Note also that the singularity is integrable at any $x$ due to the $1/\ln^{2}E$ factor.

\begin{figure}[h!]
\includegraphics[width=\linewidth]{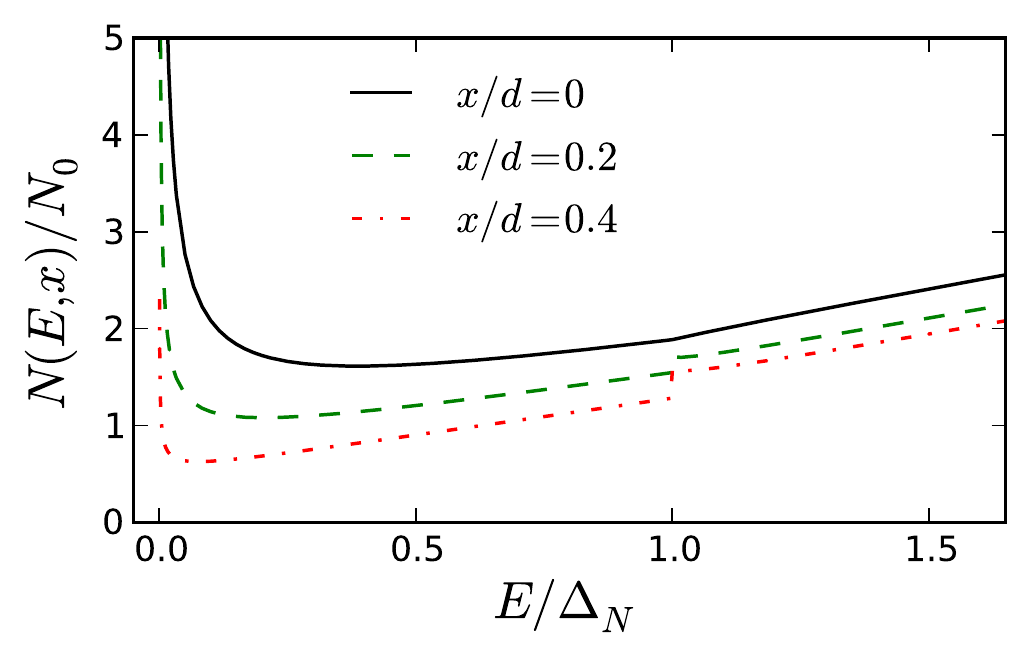}
\caption{\label{AnalyticLDOSplot} (Color online) Local density of states as a function of energy in the ballistic limit and for perfectly transmitting SN interface [Eq.~(\ref{LDOSexact})] at various positions $x$, as shown in the legend. $\Delta_Nd/v_F=0.3$.}
\end{figure}

Due to the particularly simple form of the Green's function in Eq.~(\ref{realpart}), the LDOS can be determined analytically for all energies:
\vfill
\begin{widetext}
\begin{equation} \label{LDOSexact}
N(E,x)=\frac{2\pi N_0d}{v_F}\biggl[\frac{-E+\Delta_N\cosh(\bar{\chi}\tilde{x})}{\bar{\chi}^2}\theta(\Delta_N-E)\theta(\bar{\chi}-2\Omega_Nd/v_F)+ \sum_n\frac{E-(-1)^n\Delta_N\cos(\chi_n\tilde{x})}{\chi_n^2}\theta(E-\Delta_N)\biggr].
\end{equation}
\end{widetext}
The only free parameter which enters Eq.~(\ref{LDOSexact}) is $\Delta_Nd/v_F$. In addition to controlling the amplitude of the zero-energy peak, this parameter determines the behavior of the LDOS for energies in the vicinity of $\Delta_N$.  If $2\Delta_Nd/v_F<1$, then $\bar{\chi}>2\Omega_Nd/v_F$ and the LDOS is non-zero for all energies $E<\Delta_N$. If instead $2\Delta_Nd/v_F>1$, a gap forms at energies for which $\bar{\chi}<2\Omega_Nd/v_F$. The LDOS at various positions within the normal metal is plotted in Fig.~\ref{AnalyticLDOSplot}, with $\Delta_Nd/v_F=0.3$. For all $x>0$, the LDOS is discontinuous at $E=\Delta_N$; for more details about the behavior for energies near $\Delta_N$, see Appendix~\ref{LDOSapp}.

Despite the limitations of this model, it does explicitly demonstrate the role of repulsive interactions in the normal metal. Had the pairing potential taken a positive sign in N, the pole in Eq.~(\ref{Green's function}) would have been lost and the LDOS would have exhibited a gap of size $\Delta_N$ (i.e., N becomes a superconductor). Furthermore, since the prefactor of the singularity in Eq.~(\ref{AnalyticLDOSlowenergy}) is proportional to $\Delta_N$, this term does not appear in the absence of interactions.

\subsection{Ballistic Limit with Interfacial Barrier} \label{Barrier}
We now consider the effects of a potential barrier at the SN interface. Because the quasiclassical equations described in Sec.~\ref{Quasiclassics} are valid in describing only those properties that vary slowly on the Fermi wavelength scale, the inclusion of a barrier that is sharp on the atomic scale (e.g., a delta function barrier) requires some care. Suitable boundary conditions describing barriers of this type were derived by Zaitsev \cite{Zaitsev:1984} and Kieselmann; \cite{Kieselmann:1987} they are given by
\begin{equation} \label{BC}
\begin{array}{c}
\displaystyle \hat{d}_S=\hat{d}_N, \\ \\
\displaystyle \frac{1-R(v_x)}{1+R(v_x)}\left[\hat{s}_S\left(1+\frac{1}{2}\hat{d}_S\right),\hat{s}_N\right]=-\hat{d}_S\hat{s}_S^2,
\end{array}
\end{equation}
where $R(v_x)$ is the reflection coefficient of the interface, $\hat{d}=\hat{g}(v_x,\omega,0)-\hat{g}(-v_x,\omega,0)$, and $\hat{s}=\hat{g}(v_x,\omega,0)+\hat{g}(-v_x,\omega,0)$. Because these boundary conditions produce a discontinuity in the Green's function at the interface, the normalization condition must also be explicitly imposed in the normal metal (see Ref.~\onlinecite{Ashida:1989}), $\hat{g}^2_N(v_x,\omega,x)=\hat{1}$. In the presence of a barrier, the Green's function in Eq.~(\ref{Green's function}) is modified to 
\vfill
\begin{widetext}
\begin{equation} \label{BarrierG}
g_N(v_x,\omega,x)=\frac{\omega}{\Omega_N}\frac{\Omega_N(1+R)\sinh\chi-\Delta_N(1-R)\cosh\chi+\Delta_N(1-R)\cosh(\chi\tilde{x})}{\sqrt{\bigl(\Omega_N(1+R)\cosh\chi-\Delta_N(1-R)\sinh\chi\bigr)^2-4\Omega_N^2R}}.
\end{equation}
\end{widetext}
[Equation (\ref{BarrierG}) reduces back to Eq.~(\ref{Green's function}) at $R=0$, as it should.] In order to proceed analytically, we take $R$ to be independent of $v_x$. Although $\Delta_N$ is treated as a parameter in the step model, it is to be understood that in reality $\Delta_N$ depends on $R$, as normal reflection weakens Andreev reflection and thus suppresses superconducting correlations in N; see further discussion of this effect at the end of this section. In contrast to the Green's function displayed in Eq.~(\ref{Green's function}), which contains a single pole, the Green's function in Eq.~(\ref{BarrierG}) contains branch cut singularities after analytic continuation. For energies $E<\Delta_N$, the real part of the retarded Green's function is non-zero only if $|\Omega_N(1+R)\cosh\chi-\Delta_N(1-R)\sinh\chi|<2\Omega_N\sqrt{R}$. This means that the angular integral must run over only those angles for which this inequality holds; this range of angles is given by $\zeta_1<\zeta<\zeta_2$, where we define
\begin{equation}
\begin{aligned}
\zeta_1&=\frac{2\Omega_Nd/v_F}{\sinh^{-1}\left(\frac{\Omega_N(1-R)}{|2\Delta_N\sqrt{R}-E(1+R)|}\right)}, \\
\zeta_2&=\frac{2\Omega_Nd/v_F}{\sinh^{-1}\left(\frac{\Omega_N(1-R)}{2\Delta_N\sqrt{R}+E(1+R)}\right)}.
\end{aligned}
\end{equation}
We can immediately make two qualitative conclusions about the LDOS. First, due to terms that behave as $\exp(1/\zeta)$, the angular average of the retarded Green's function will diverge if the lower limit of integration ($\zeta_1$) goes to zero. This will produce a singularity in the LDOS at an energy
\begin{equation} \label{newpeak}
E_0=\frac{2\Delta_N\sqrt{R}}{1+R},
\end{equation}
which is the energy at which $\zeta_1\to0$ (this result is also found in Ref.~\onlinecite{Belzig:2000}). Second, a gap will form in this system at energies for which $\zeta_1>1$. We now investigate these two properties of the system further.

\begin{figure}
\includegraphics[width=\linewidth]{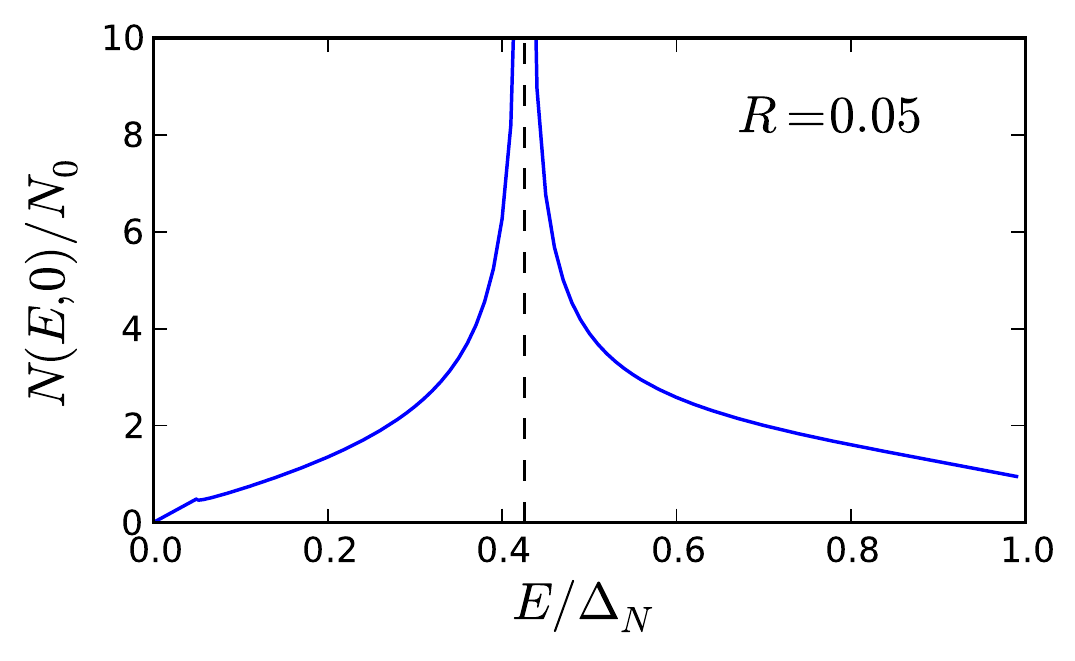}
\caption{\label{BarrierFig} (Color online) Local density of states as a function of energy in the ballistic limit with a non-ideal SN interface parameterized by reflection coefficient $R=0.05$. $\Delta_Nd/v_F=0.4$; vertical dashed line corresponds to $E_0$, as given by Eq.~(\ref{newpeak}).}
\end{figure}

To determine the form of the singularity, we expand for $E=E_0+\delta E$, assuming that $|\delta E|$ is much smaller than both $E_0$ and $\Delta_N(1-R)$. For $\zeta_1$, we obtain
\begin{equation}
\zeta_1=\frac{2\Delta_Nd}{v_F}\frac{1-R}{1+R}\frac{1}{\ln\left(2\frac{(1-R)^2}{(1+R)^2}\frac{\Delta_N}{|\delta E|}\right)},
\end{equation}
and, defining $\chi'=\frac{2\Delta_Nd}{v_x}\frac{1-R}{1+R}$, the LDOS is given by
\begin{equation}
N(E,x)=\frac{1+R}{1-R}\int_{\zeta_1}^{\text{min}\{\zeta_2,1\}}d\zeta\frac{\cosh(\chi'\tilde{x})-e^{-\chi'}}{\sqrt{1-\frac{(1+R)^2}{4R}e^{-2\chi'}}}.
\end{equation}
Provided that $(\Delta_N/|\delta E|)^{\tilde{x}}\gg1$, we can further expand
\begin{equation}
N(E,x)=\frac{1+R}{2(1-R)}\int_{\zeta_1}^{\text{min}\{\zeta_2,1\}}d\zeta\,\exp(\chi'\tilde{x}).
\end{equation}
This integral is dominated by the contribution from the lower limit and can be approximated by
\begin{equation} \label{LDOSBarrier}
\begin{aligned}
N(E,x)&=\frac{N_0\Delta_Nd}{v_F\tilde{x}}\frac{1}{\ln^2\left(2\frac{(1-R)^2}{(1+R)^2}\frac{\Delta_N}{|\delta E|)}\right)}\left(2\frac{(1-R)^2}{(1+R)^2}\frac{\Delta_N}{|\delta E|}\right)^{\tilde{x}}.
\end{aligned}
\end{equation}
While the presence of the barrier shifts the LDOS singularity from $E=0$ to $E=E_0$, the functional form of the singularity remains unchanged. A sample plot of the density of states for $R=0.05$ and $\Delta_Nd/v_F=0.4$ is shown in Fig.~\ref{BarrierFig}. This plot also demonstrates an important property of Eq.~(\ref{newpeak}): even a small amount of normal reflection at the SN interface produces a substantial shift in the position of the LDOS singularity (compared to $\Delta_N$) since $E_0/\Delta_N\propto \sqrt{R}$ rather than to $R$ itself.

For certain choices of the parameters $\Delta_Nd/v_F$ and $R$, it is also possible to have a mini-gap in the LDOS. This occurs if $\zeta_1>1$ and thus the range of integration over $\zeta$ shrinks to zero. In particular, a gap is formed around $E=0$ if $R>\exp(-4\Delta_Nd/v_F)$. For parameter values $\Delta_Nd/v_F\lesssim1$, the size of this gap can be well approximated by
\begin{equation}
E_g=\frac{\sqrt{R}}{1-R}\ln(1/R)\left(\Delta_N-\frac{v_F\ln(1/R)}{4d}\right).
\end{equation}
This is consistent with the limiting case $R\to1$, in which case a gap of size $\Delta_N$ is formed around the Fermi energy. Additionally, a gap is formed around $E=\Delta_N$ if $\sqrt{R}<\frac{2\Delta_Nd/v_F-1}{2\Delta_Nd/v_F+1}$, which is consistent with the limiting case $R\to0$. As shown in Sec.~\ref{Ballistic}, a gap forms around $\Delta_N$ if $2\Delta_Nd/v_F>1$ in this limit.

While the step model is useful for describing some of the qualitative features of the LDOS in the presence of the barrier, it falls short quantitatively. In particular, $\Delta_N$ is treated as a model parameter here, while in reality it should also depend on $R$. For constant $R$, the spatial profile of the induced pairing potential is modified to $\Delta(x)\sim-\lambda_N N_0v_F/x\times\frac{1-R}{1+R}$ for $x\gg\xi_S$, so that it is suppressed as $R\to1$. Also, in reality $R$ is not independent of $v_x$. For example, the reflection coefficient of an interface potential $V(x)=V_0\delta(x)$ is given by $R(v_x)=V_0^2/(V_0^2+v_x^2)$. Including these two effects would require a fully self-consistent numerical solution in the presence of the barrier, so we simply note the qualitative features induced by the barrier and proceed throughout the rest of the paper under the assumption that there is no interface potential.

\subsection{Diffusive Limit} \label{Diffusive}
While it is not possible to solve Eq.~(\ref{Eilenberger}) analytically for arbitrary impurity concentration, it is possible to solve it analytically in the diffusive limit. In this limit, the Eilenberger equation reduces to the Usadel equation, \cite{Usadel:1970} which in a region of constant $\Delta$ reads
\begin{equation} \label{Usadel}
\frac{d^2\theta}{dx^2}=\frac{2\Omega}{D}\sin\bigl(\theta(\omega,x)-\delta\bigr).
\end{equation}
In this parameterization, $D=v_F\ell/3$ is the diffusion coefficient, $\delta_S=\tan^{-1}(\Delta_S/\omega)$, and $\delta_N=-\tan^{-1}(\Delta_N/\omega)$. The Green's functions are expressed through the function $\theta(\omega,x)$ as $\sin\theta(\omega,x)=\avg{f(v_x,\omega,x)}$ and $\cos\theta(\omega,x)=\avg{g(v_x,\omega,x)}$. Equation (\ref{Usadel}) was solved in Ref.~\onlinecite{Altland:1998} for the non-interacting case ($\Delta_N=0$); the solution with repulsive interactions is obtained along the same lines as
\vfill
\begin{widetext}
\begin{equation} \label{Usadelsolution}
\begin{array}{c}
\displaystyle \theta_N(x)=\delta_N+2\sin^{-1}\biggl[\sin(\alpha_1/2)\text{sn}\biggl(-i\sqrt{\frac{2\Omega_N}{D}}(d-x)+K\bigl[\sin^2(\alpha_1/2)\bigr]\biggr|\sin^2(\alpha_1/2)\biggr)\biggr], \\ \\
\displaystyle	\theta_S(x)=\delta_S+4\tan^{-1}\biggl[\exp\biggl(\sqrt{\frac{2\Omega_S}{D}}x\biggr)\tan\biggl(\frac{\alpha_0+\delta_N-\delta_S}{4}\biggr)\biggr],
\end{array}
\end{equation}
\end{widetext}
where $\alpha_0=\theta_N(\omega,0)-\delta_N$, $\alpha_1=\theta_N(\omega,d)-\delta_N$, $\text{sn}(u|m)$ is a Jacobi elliptic function, and $K(m)$ is the complete elliptic integral of the first kind (see Ref.~\onlinecite{Abramowitz}). The remaining boundary values $\alpha_0$ and $\alpha_1$ must be determined numerically by requiring continuity of $\theta(\omega,x)$ and $d\theta/dx$ across the SN interface.

\begin{figure} [t!]
\includegraphics[width=\linewidth]{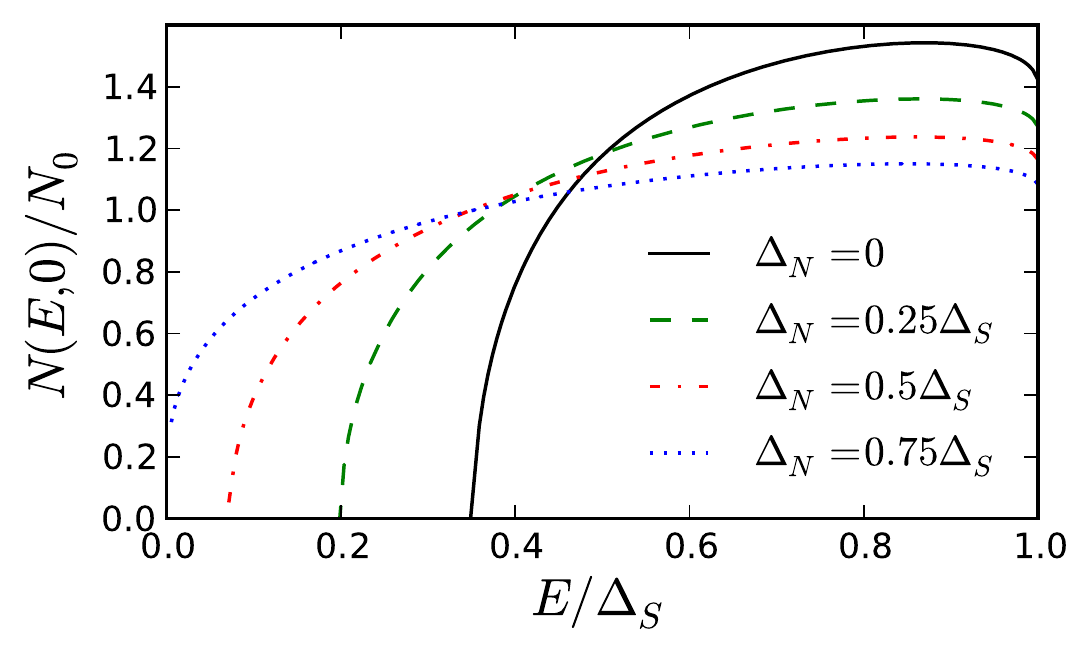}
\caption{ \label{UsadelDOS} (Color online) Local density of states in the diffusive limit at the superconductor/normal metal interface ($x=0$) plotted for various values of $\Delta_N$. The normal metal thickness is $d=\sqrt{D/2\Delta_S}$. 
}
\end{figure}

It is well known that, in the absence of interactions, a uniform mini-gap of size $E_g\sim D/d^2$ forms throughout the entire normal metal. \cite{McMillan:1968,Golubov:1988,Belzig:1996} We find similar results in the presence of repulsive interactions, whereby the interactions simply reduce the size of the the mini-gap. For sufficiently large values of $\Delta_N$, the system becomes gapless. Nevertheless, a zero-energy peak does not occur. The LDOS in the diffusive limit is plotted in Fig. \ref{UsadelDOS} for several values of $\Delta_N$ in a system with $d=\sqrt{D/2\Delta_S}$.

\section{\label{Numerical} 
Arbitrarily Strong Disorder}
In this section, we investigate how the zero-energy singularity in the LDOS transforms into a hard mini-gap with increasing disorder. For arbitrary disorder, i.e., when we are neither in the ballistic limit nor in the diffusive limit, Eq.~(\ref{Eilenberger}) must be solved self-consistently using numerical methods. We thus performed self-consistent calculations of both the pairing potential $\Delta(x)$ and the self-energy $\hat{\sigma}(\omega,x)$, from which we obtain the Green's function and, eventually, the LDOS. 

\subsection{\label{Methods} Numerical Methods}
For our numerical calculations, we use the Riccati parameterization of the Eilenberger equations. In this parameterization, Eq.~(\ref{Eilenberger}) decouples into two stable first-order equations of the Riccati type via a transformation of the form \cite{Schopohl:1998}
\begin{equation} \label{Schopohl}
a=\frac{f(v_x,\omega,x)}{1+g(v_x,\omega,x)},\hspace*{0.1in}b=\frac{f^\dagger(v_x,\omega,x)}{1+g(v_x,\omega,x)}.
\end{equation}
In terms of these new functions, the Green's function is parameterized by
\begin{equation} \label{Gparam}
\hat{g}(v_x,\omega,x)=\frac{1}{1+ab}\left(\begin{array}{cc} 1-ab & 2a \\ 2b & ab-1 \end{array}\right).
\end{equation}
After applying transformation (\ref{Schopohl}), the resulting Riccati differential equations are
\begin{equation} \label{Riccati}
\begin{aligned}
-v_x\partial_xa&=\bigl(a^2-1\bigr)\tilde{\Delta}(x)+2\tilde{\omega}(x)a, \\
v_x\partial_xb&=\bigl(b^2-1\bigr)\tilde{\Delta}(x)+2\tilde{\omega}(x)b,
\end{aligned}
\end{equation}
where the impurity self-energy is included in the definitions $\tilde{\omega}(x)=\omega+\sigma_{11}(\omega,x)$ and $\tilde{\Delta}(x)=\Delta(x)+\sigma_{12}(\omega,x)$. When working with real 
rather than Matsubara 
frequencies, $\omega$ is simply replaced by $-iE$ in these definitions. Equations (\ref{Riccati}) can be integrated in the stable direction using the expression \cite{Pilgram:2000}
\begin{equation} \label{Riccatiint}
a_{n+1}=\frac{[\tilde{\Delta}-(\tilde{\Omega}+\tilde{\omega})a_n]e^{-2\tilde{\Omega} h/v_x}-\tilde{\Delta}-(\tilde{\Omega}-\tilde{\omega})a_n}{[\tilde{\Delta}a_n-\tilde{\Omega}+\tilde{\omega}]e^{-2\tilde{\Omega}h/v_x}-\tilde{\Delta}a_n-\tilde{\Omega}-\tilde{\omega}},
\end{equation}
where $\tilde{\Omega}=(\tilde{\Delta}^2+\tilde{\omega}^2)^{1/2}$ and $h$ is the step size.

Because determining $\Delta(x)$ requires Matsubara frequencies while determining $N(E,x)$ requires real frequencies, we perform these two calculations in parallel. We begin by solving Eqs. (\ref{Riccati}) in the clean, non-interacting limit (i.e., $\Delta(x)=\Delta_S\theta(-x)$ and $\hat{\sigma}(\omega,x)=\hat{\sigma}^R(E,x)=0$), using both real and Matsubara frequencies. We then construct the retarded and Matsubara Green's functions through Eq.~(\ref{Gparam}) and substitute these Green's functions into Eqs. (\ref{delta}) and (\ref{selfenergy}) to obtain $\Delta(x)$, $\hat{\sigma}(\omega,x)$, and $\hat{\sigma}^R(E,x)$. Because the retarded Green's functions in the clean, non-interacting limit contain poles, we perform these angular averages analytically; for more details, see Appendix~\ref{mini-gap}. Finally, we substitute the calculated functions $\Delta(x)$, $\hat{\sigma}(\omega,x)$, and $\hat{\sigma}^R(E,x)$ back into Eqs. (\ref{Riccati}) and iterate numerically until self-consistency is achieved.

\begin{figure}[t!] 
\includegraphics[width=\linewidth]{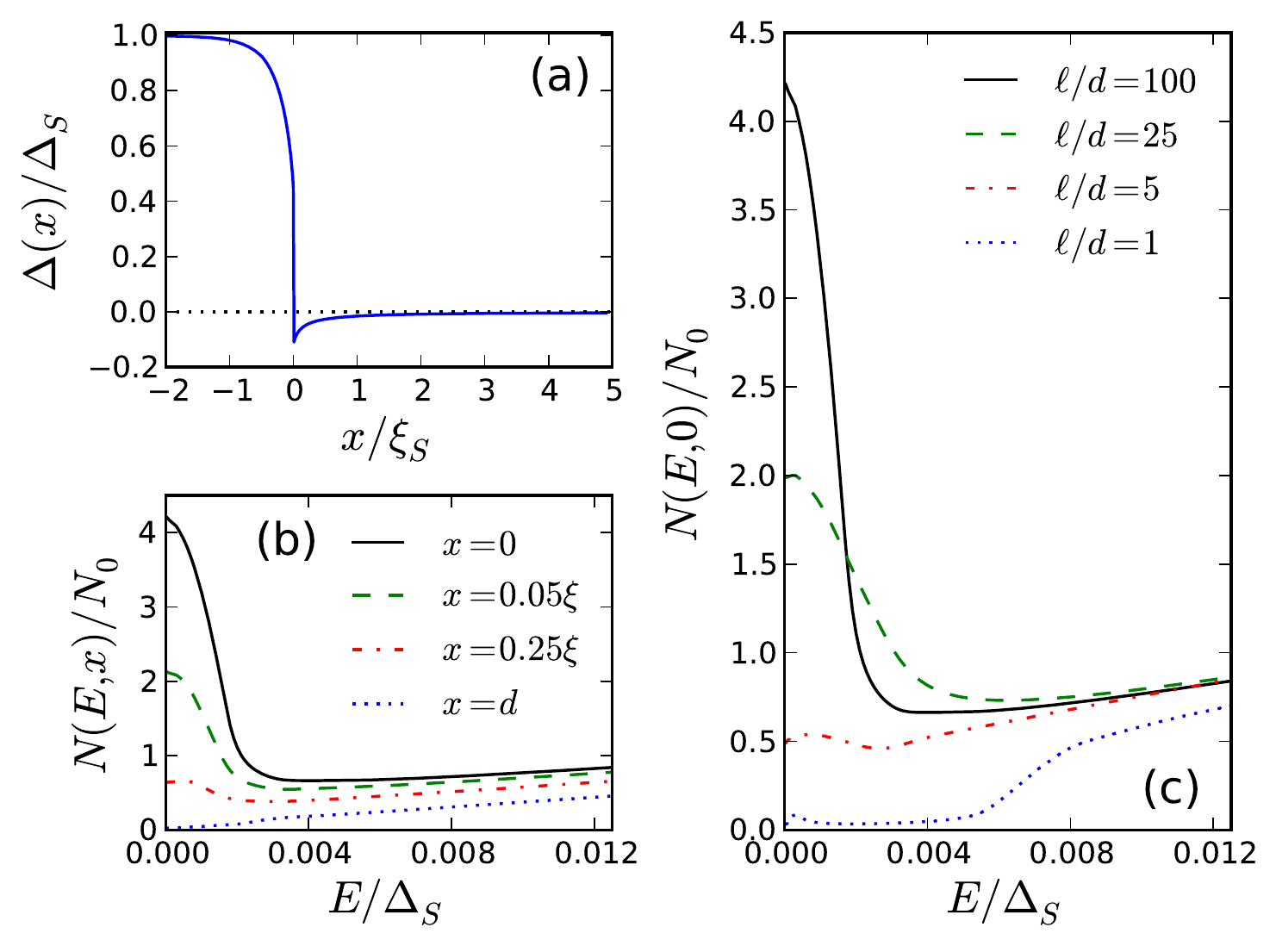}
\caption{\label{numplot} (Color online) (a) Spatial profile of the pairing potential $\Delta(x)$ following from a self-consistent solution of Eqs.~(\ref{Eilenberger})--(\ref{selfenergy}) with coupling constants $\lambda_SN_0=1$ and $\lambda_NN_0=-0.25$ and normal metal thickness $d=10\xi_S$. (b) Energy dependence of LDOS at fixed mean free path $\ell=100d$, shown for various values of $x$. (c) Energy dependence of LDOS at fixed position $x=0$, plotted for various values of mean free path $\ell$.}
\end{figure}

\subsection{\label{Results} Results}
Figure~\ref{numplot} summarizes the results of our self-consistent calculation. Figure~\ref{numplot}(a) shows the spatial profile of the pairing potential $\Delta(x)$, calculated with coupling constants $\lambda_SN_0=1$ and $\lambda_NN_0=-0.25$ and normal metal thickness $d=10\xi_S$. The profile is largely unaffected by disorder and is shown for $\ell/d=100$.

In the absence of interactions, disorder opens up a mini-gap in the normal metal even in the quasi-ballistic limit, i.e., for infinitesimally small values of $1/\ell$ (Ref.~\onlinecite{Pilgram:2000}; see also Appendix~\ref{mini-gap}). We find that in the presence of repulsive interactions, the quasi-ballistic mini-gap is eliminated and the zero-energy peak in the LDOS persists to finite values of $\ell$. Figure~\ref{numplot}(b) shows the LDOS at various positions  within the normal metal. The peak is most pronounced at the SN interface and is localized to the interface on a scale $\xi_S$.

Figure~\ref{numplot}(c) shows the evolution of the zero-energy peak with decreasing mean free path. We find that the amplitude of the peak is very sensitive to disorder. The peak remains distinct down to $\ell/d\sim10$, while at $\ell/d=5$ the peak is far less discernible. For $\ell\sim d$ the peak is strongly suppressed and the mini-gap, shown in Sec.~\ref{Step}~\ref{Diffusive} to occur in the limit $\ell\ll d$, starts to develop.

\section{\label{FiniteT} Finite Temperature}
In this section, we study the sensitivity of the zero-energy LDOS peak to finite temperature. At finite temperature, the self-consistency condition on the pairing potential becomes
\begin{equation} \label{Tself-consistent}
\Delta(x)=-\pi\lambda(x)N_0T\sum_{\omega_n}\avg{f(v_x,\omega_n,x)},
\end{equation}
where $\omega_n=(2n+1)\pi T$ are the fermionic Matsubara frequencies.
For $T\ll T_c$, there are now two spatial scales in the problem: the superconducting coherence length, $\xi_S$, and the normal metal coherence length, $\xi_T=v_F/T\gg \xi_S $  in the ballistic limit. The $1/x$ decay of the pairing potential is cut off at $\xi_T$; for $x\gg \xi_T$, when only the first Matsubara frequency is important, the pairing potential falls of exponentially as
\begin{equation}
\Delta(x)\sim-(\lambda_NN_0v_F/x)e^{-2\pi x/\xi_T}.
\end{equation}

\begin{figure}[t!]
\includegraphics[width=\linewidth]{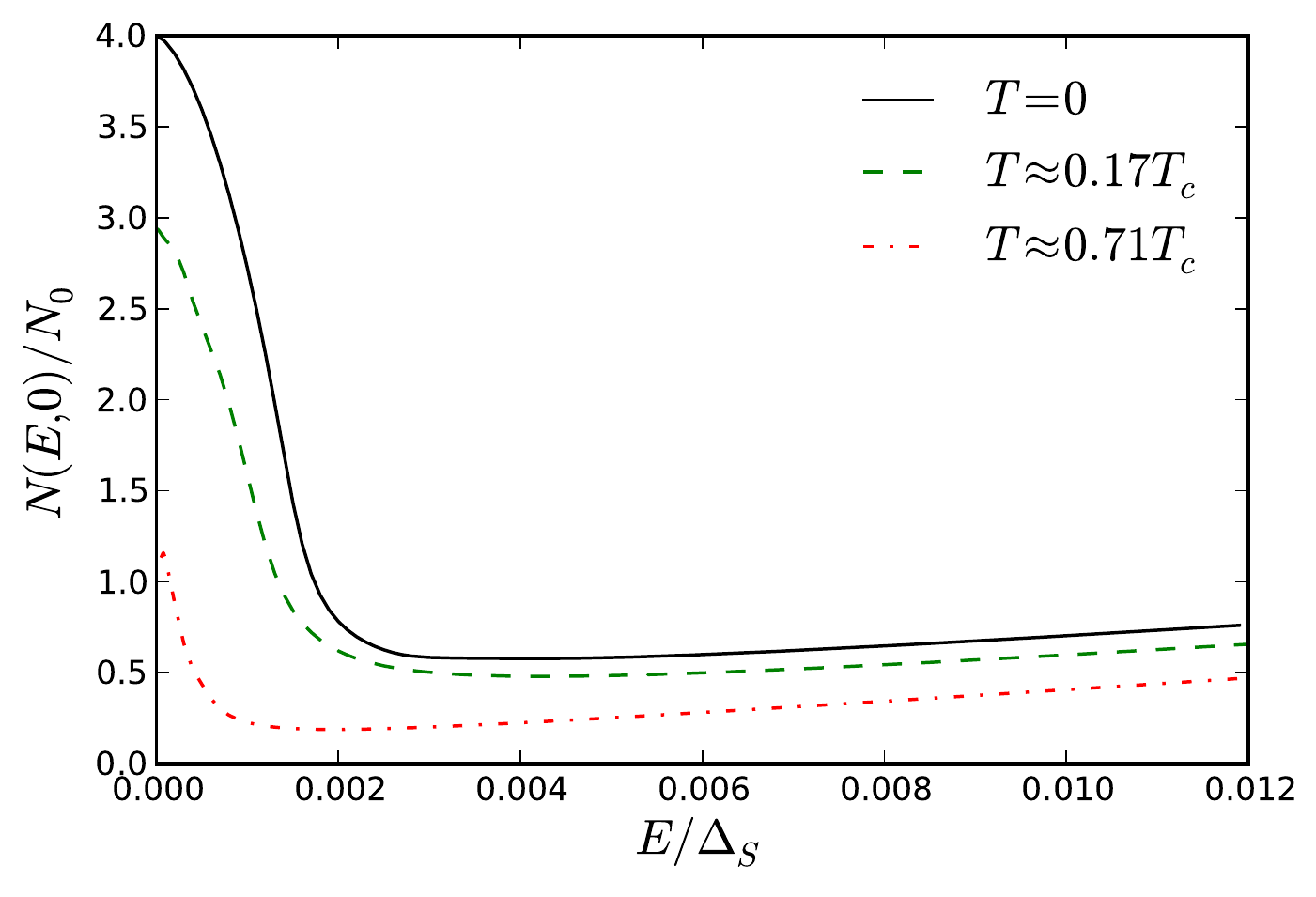}
\caption{\label{temperature} (Color online) Energy dependence of LDOS at fixed position $x=0$ and mean free path $\ell=100d$, plotted at several different temperatures. LDOS follows from a self-consistent calculation of $\Delta(x)$ using Eq.~(\ref{Tself-consistent}); we choose $\lambda_SN_0=0.5$ and $\lambda_NN_0=-0.25$.}
\end{figure}

We self-consistently calculated both the pairing potential (choosing $\lambda_SN_0=0.5$ and $\lambda_NN_0=-0.25$) and the LDOS in the quasi-ballistic limit ($\ell=100d$); the results are displayed in Fig.~\ref{temperature}. Unlike with disorder, we find that the zero-energy peak is rather robust to finite temperatures. At $T\approx0.17T_c$ there is a distinct zero-energy peak, while even at $T\approx 0.71T_c$ the zero-energy peak is not completely smeared. However, as the temperature is increased, the zero-energy peak becomes more localized to the SN interface.

\section{\label{Beyond QC} Beyond Quasiclassics}
Recently, an SN junction with repulsive interactions on the N side was studied numerically by using the exact (non-quasiclassical) BdG equation. \cite{Valls:2010} This study found no evidence of a zero-energy peak in the LDOS. The authors of Ref.~\onlinecite{Valls:2010} asserted that this peak is an artifact of the quasiclassical approximation. To test this assertion, we performed our own non-quasiclassical calculation.

\begin{figure}[t!]
\includegraphics[width=\linewidth]{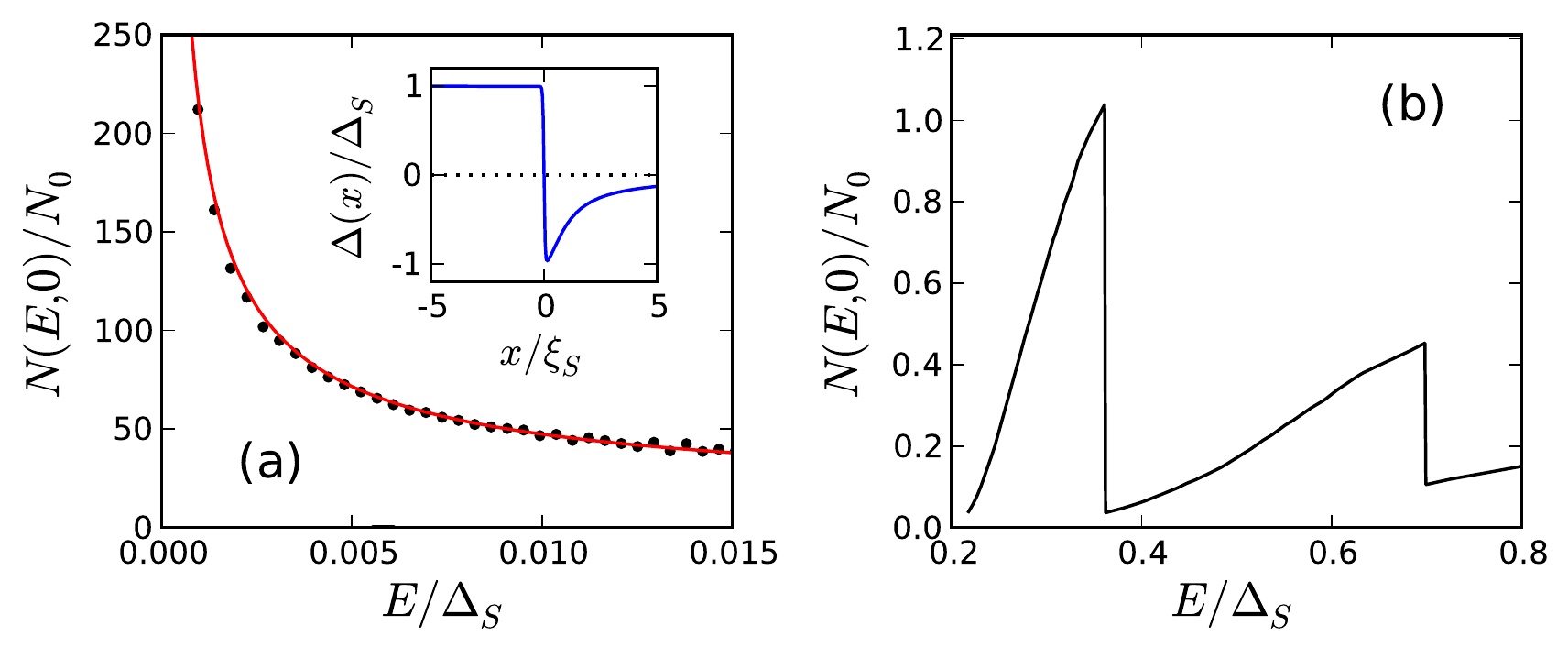}
\caption{\label{BdGDOS} (Color online) Local density of states at SN interface, calculated using exact (non-quasiclassical) BdG equation [Eq.~(\ref{BdG})] with $d=5\xi_S$ and $\mu=100\Delta_S$. (a) Zoom of low-energy behavior. Both the numerical results (dots) and $1/E\ln^2E$ functional form predicted by Eq.~(\ref{AnalyticLDOSlowenergy}) (solid line) are plotted. Inset: Model form of pairing potential $\Delta(x)$. (b) Andreev structure of LDOS for energies $E\gtrsim v_F/d$.}
\end{figure}

We solve the BdG equation [Eq.~(\ref{BdG})] in a finite-sized system via numerical diagonalization, using a finite difference method to approximate the derivative and choosing $d=5\xi_S$ and $\mu=100\Delta_S$. Since our primary goal is to demonstrate the existence of the zero-energy bound state beyond the quasiclassical approximation rather than study it in detail, we limit ourselves to a non-self-consistent calculation for a suitable choice of the pairing potential $\Delta(x)$. In order to emphasize the zero-energy peak, we choose a model form of $\Delta(x)$ that exaggerates the repulsive interaction on the N side [Fig.~\ref{BdGDOS}(a) inset]. The LDOS (at positive energies only) is calculated from the BdG wave functions through
\begin{equation}
N(E,x)=\sum_n\int\frac{d^2k_\perp}{(2\pi)^2}|u_n(k_\perp,x)|^2\delta[E-E_n(k_\perp)].
\end{equation}
At a given value of $k_\perp$, the wave functions are normalized according to
\begin{equation}
\int dx\biggl[|u_n(k_\perp,x)|^2+|v_n(k_\perp,x)|^2\biggr]=1.
\end{equation}
The results of the numerical calculation are displayed in Fig.~\ref{BdGDOS}. We find a distinct peak in the LDOS at low energies [Fig.~\ref{BdGDOS}(a)] superimposed on the usual Andreev structure  \cite{deGennes:1963} that exists for energies $E\gtrsim v_F/d$ [Fig.~\ref{BdGDOS}(b)]. Not only does our numerical calculation show a zero-energy peak, but this peak reproduces the $1/E\ln^2E$ form predicted by Eq.~(\ref{AnalyticLDOSlowenergy}).

\begin{figure}[t!]
\includegraphics[width=\linewidth]{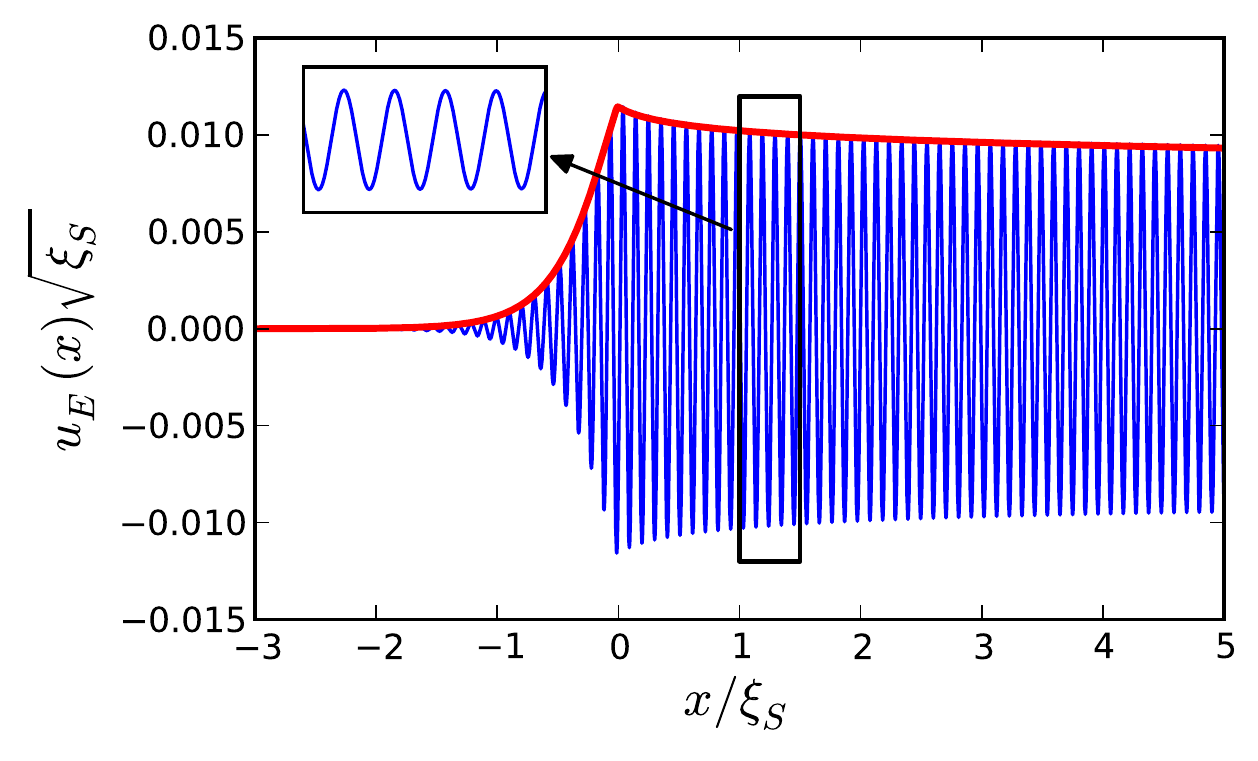}
\caption{\label{BdGplot} (Color online) Spatial dependence of the (unnormalized) wave function $u_E(x)$ corresponding to the minimal eigenvalue $E$ of the BdG equation (rapidly oscillating curve). Quasiclassical wave function [Eq.~(\ref{BdGsolution})] traces the envelope of the exact wave function (slowly varying curve). Parameters: $\mu=100\Delta_S$ and $(1-k_\perp^2/k_F^2)^{1/2}=v_x/v_F=0.3$. For $\Delta(x)$, we use the form shown in Fig.~\ref{numplot}(a). Inset: Zoom of oscillations on the Fermi wavelength scale.}
\end{figure}

We find further evidence supporting the validity of the quasiclassical approximation by simply examining the wave functions themselves. As discussed in Sec.~\ref{Intro}, the wave function corresponding to the bound state [see Eq.~(\ref{BdGsolution})] should oscillate on the scale of the Fermi wavelength with an envelope that decays as a power law into the normal metal and as an exponential into the superconductor. Figure~\ref{BdGplot}(a) shows the spatial dependence of the wave function $u_E(x)$ corresponding to the minimal eigenvalue of Eq.~(\ref{BdG}) (the rapidly oscillating curve), calculated using the profile of $\Delta(x)$ shown in Fig.~\ref{numplot}(a) and with $\mu=100\Delta_S$ and $(1-k_\perp^2/k_F^2)^{1/2}=0.3$. We also plot the upper component of the quasiclassical BdG spinor $\hat{\phi}_0(x)$ given in Eq.~(\ref{BdGsolution}) (the slowly varying curve) with $v_x/v_F=0.3$, which traces the envelope of the exact BdG wave function.

Based both on the zero-energy LDOS peak that we obtain by solving the full BdG equation as well as the accuracy of Eq.~(\ref{BdGsolution}) in describing the envelope of the exact wave function, we conclude that, contrary to the assertion of Ref.~\onlinecite{Valls:2010}, the quasiclassical approximation does a very good job in describing an SN junction.

\section{Conclusions} \label{Conclusions}
We considered the effects of impurity scattering and a tunnel barrier on the zero-energy bound state that forms at the interface between a conventional $s$-wave superconductor and a normal metal with repulsive electron-electron interactions. We showed, through a combination of analytical and numerical calculations, that disorder weakens the zero-energy peak in the local density of states to the point that a mini-gap develops in the diffusive limit. Furthermore, an interfacial barrier shifts this zero-energy peak to a finite energy. Additionally, we went beyond the quasiclassical approximation to show numerically that the zero-energy bound state can be obtained through non-quasiclassical means as well. Based on the results of this paper, we conclude that the zero-energy local density of states peak relies strongly on both the good quality of the sample and the SN interface.

As we said in Sec.~\ref{Intro}, there has been no direct experimental observation of the zero-energy peak. Our study reveals one possible reason for the lack of experimental evidence: the zero-energy peak is very sensitive to disorder and can be seen only in nearly ballistic normal metal films. On the other hand, SN junctions with N being a conventional non-superconducting metal (silver, gold, etc.) are typically highly disordered.  One possible solution would be to replace N by a high-mobility semiconductor heterostructure; however the electron (or hole) layers in these devices are buried under the insulating cap and thus not accessible to scanning tunneling microscopy (STM). We propose to search for the zero-energy peak in suspended graphene, which has quite high electron mobilities $\sim10^5$ cm$^2$ V$^{-1}$ s$^{-1}$ while also offering an exposed two-dimensional surface. Furthermore, graphene likely has repulsive interactions because it shows no tendency toward intrinsic superconductivity. To focus on the physics discussed in this paper, one needs to back-gate the Fermi energy away from the Dirac point. Independent experiments have demonstrated the feasibility of performing STM, \cite{Zan:2012} achieving ballistic transport, \cite{Bolotin:2008} and inducing the proximity effect \cite{Mizuno:2013} in this material; these are the three main criteria needed to observe the zero-energy peak.
 
\acknowledgements
We thank C. Beenakker, W. Belzig, G. Blatter, C. Bruder, P. Goldbart, M. Graf, S. Lin, S. Maiti, O. Millo, Y. Tanaka, S. Tessmer, and V. Zyuzin for useful discussions.
This work was supported by the National Science Foundation via Grant No. DMR-1308972. D.L.M acknowledges hospitality of the Aspen Center of Physics where a part of the work was done.  

\appendix
\section{LDOS near $E=\Delta_N$} \label{LDOSapp}

In this appendix, we discuss in further detail the behavior of the LDOS at energies near $\Delta_N$ in the step potential model. We begin with the result from Sec.~\ref{Ballistic},
\vfill
\begin{widetext}
\begin{equation} \label{DOSapp}
N(E,x)=\frac{2\pi N_0d}{v_F}\biggl[\frac{-E+\Delta_N\cosh(\bar{\chi}\tilde{x})}{\bar{\chi}^2}\theta(\Delta_N-E)\theta(\bar{\chi}-2\Omega_Nd/v_F)+ \sum_n\frac{E-(-1)^n\Delta_N\cos(\chi_n\tilde{x})}{\chi_n^2}\theta(E-\Delta_N)\biggr],
\end{equation}
\end{widetext}
where we defined $\bar{\chi}=\tanh^{-1}(\Omega_N/\Delta_N)$ and $\chi_n=\tan^{-1}(\Omega_N/\Delta_N)+n\pi$, and the sum runs over all $n>(2\Omega_Nd/v_F-\chi_0)/\pi$.

We first consider the case $2\Delta_Nd/v_F<1$, so that the LDOS is non-zero for all energies $E<\Delta_N$ and the sum in Eq.~(\ref{DOSapp}) starts at $n=0$. Expanding Eq.~(\ref{DOSapp}) for $E=\Delta_N-\delta E$, with $0<\delta E\ll\Delta_N$, gives
\begin{equation} \label{Eless}
N(E,x)=\frac{\pi N_0\Delta_Nd}{v_F}\left[(1+\tilde{x}^2)+\frac{1}{6}(-5+\tilde{x}^4)\frac{\delta E}{\Delta_N}\right].
\end{equation}
For energies $E>\Delta_N$, it is illustrative to separate the $n=0$ term in the sum,
\begin{equation} \label{Egreater}
\begin{aligned}
N(E,x)&=\frac{2\pi N_0d}{v_F}\biggl(\frac{E-\Delta_N\cos(\chi_0\tilde{x})}{\chi_0^2}+ \\
	&+ \sum_{n=1}^\infty\frac{E-(-1)^n\Delta_N\cos(\chi_n\tilde{x})}{\chi_n^2}\biggr).
\end{aligned}
\end{equation}
Expanding the $n=0$ term for $E=\Delta_N+\delta E$ gives
\begin{equation}
\frac{E-\Delta_N\cos(\chi_0\tilde{x})}{\chi_0^2}=\frac{\Delta_N}{2}(1+\tilde{x}^2)+\frac{\delta E}{12}(5-\tilde{x}^4).
\end{equation}
The LDOS contribution from the $n=0$ term for energies $E>\Delta_N$ matches the LDOS for energies $E<\Delta_N$ given in Eq.~(\ref{Eless}). Therefore, any discontinuities in the LDOS at $E=\Delta_N$ will arise from the terms corresponding to $n>0$. At $x=0$, this contribution can be expanded as
\begin{equation} \label{lincont}
\frac{2\pi N_0d}{v_F}\sum_{n=1}^\infty\frac{E-\Delta_N\cos\chi_0}{\chi_n^2}=\frac{2\pi N_0d}{3v_F}\delta E.
\end{equation}
Therefore, at $x=0$ the LDOS is given by
\begin{equation}
\begin{aligned}
N(E<\Delta_N,0)&=\frac{2\pi N_0d}{v_F}\left[\Delta_N-\frac{1}{3}(\Delta_N-E)\right], \\
N(E>\Delta_N,0)&=\frac{2\pi N_0d}{v_F}\left[\Delta_N+\frac{2}{3}(E-\Delta_N)\right].
\end{aligned}
\end{equation}
We see that the LDOS itself is continuous at $E=\Delta_N$ but experiences a kink because the slope is discontinuous there. If $x>0$, however, the contribution from terms corresponding to $n>0$ in Eq.~(\ref{Egreater}) is very different. In this case, we expand
\begin{equation} \label{sqrtcont}
\begin{aligned}
\frac{2\pi N_0d}{v_F}&\sum_{n=1}^\infty\frac{E-(-1)^n\Delta_N\cos(\chi_n\tilde{x})}{\chi_n^2}=\frac{2\pi N_0\Delta_Nd}{v_F}\times \\
	&\times\sum_{n=1}^\infty\biggl(\frac{1-(-1)^n\cos(n\pi\tilde{x})}{n^2\pi^2}-\frac{\sqrt{2\delta E/\Delta_N}}{n^3\pi^3}\times \\
	&\times\left\{2-(-1)^n\bigl[2\cos(n\pi\tilde{x})+n\pi\tilde{x}\sin(n\pi\tilde{x})\bigr]\right\}\biggr).
\end{aligned}
\end{equation}
The first term in Eq.~(\ref{sqrtcont}) produces a discontinuity in the LDOS at $E=\Delta_N$, while the second term determines that the LDOS behaves as $N(E,x)\sim-N_0\sqrt{\delta E/\Delta_N}$ rather than linearly with $\delta E$ for energies $E>\Delta_N$. Both cases $x=0$ and $x>0$ are displayed clearly in Fig. \ref{AnalyticLDOSplot}.

If instead $2\Delta_Nd/v_F<1$, then a gap forms at energies $E<\Delta_N$ and the sum in Eq.~(\ref{DOSapp}) begins at $n=1$. Therefore, the LDOS at energies $E>\Delta_N$ is given by Eq.~(\ref{lincont}) and Eq.~(\ref{sqrtcont}) for the cases $x=0$ and $x>0$, respectively.

\section{\label{mini-gap} Mini-gap in a non-interacting SN junction in the quasi-ballistic limit}
Pilgram \emph{et al.} \cite{Pilgram:2000} studied numerically the formation of a minigap in a noninteracting SN junction for arbitrary values of the ratio $d/\ell$, where $d$ is the thickness of the normal film. Their results suggest that the minigap is present for any finite $d/\ell$ and its magnitude is on the order of $1/\tau=v_F/\ell$ in the ballistic limit.\cite{Beenakker:2005}  Thus, the formation of a mini-gap is a non-perturbative effect which must be signaled by a breakdown of the perturbation theory in disorder. In this appendix, we demonstrate how this breakdown occurs.

To calculate the correction to the LDOS to first order in $1/\ell$, we solve the Riccati equations with a non-interacting step model for the pairing potential, $\Delta(x)=\Delta_S\theta(-x)$. In the Riccati parameterization [Eq.~(\ref{Gparam})], the normal Green's function can be expanded to first order in $1/\ell$ as
\begin{equation} \label{gexp}
g=\frac{1-ab}{1+ab}\approx\frac{1-a_0b_0}{1+a_0b_0}-\frac{2(a_0\delta b+b_0\delta a)}{(1+a_0b_0)^2},
\end{equation}
where $a_0$ and $b_0$ denote the solutions to Eqs. (\ref{Riccati}) in the absence of disorder and $\delta a$ and $\delta b$ denote the first-order corrections. Here and in the following, arguments of all functions are dropped for brevity and all quantities are dimensionless. All energies are given in units of $\Delta_S$, all lengths in units of $\xi_S$, and the LDOS in units of $N_0$ (the LDOS of a normal metal). Because the mini-gap appears only in the limit $E\to0$ for a system with $1/\ell\to0$, we expand all quantities in the low-energy limit.

Let us first review some of the results for a ballistic system. The solutions to Eqs. (\ref{Riccati}) in the normal metal are (assuming $\zeta>0$)
\begin{equation}
\begin{aligned}
a_0&=(\Omega_S-\omega)e^{-2\omega x/\zeta}, \\
b_0&=(\Omega_S-\omega)e^{-2\omega(2d-x)/\zeta}.
\end{aligned}
\end{equation}
Given these solutions, the Green's functions in the normal metal can be constructed as
\begin{equation}
\begin{aligned}
g_N&=\frac{\omega\cosh\chi+\Omega_S\sinh\chi}{\Omega_S\cosh\chi+\omega\sinh\chi}, \\
f_N&=\frac{e^{2\omega(d-x)/\zeta}}{\Omega_S\cosh\chi+\omega\sinh\chi},
\end{aligned}
\end{equation}
where in the absence of interactions $\chi=2\omega d/|\zeta|$. After analytic continuation, these Green's functions contain poles located at $\chi_n=\tan^{-1}(\Omega_S/E)+n\pi$. The self-energy is calculated from the retarded Green's functions:
\begin{equation} \label{self-energycalc}
\begin{aligned}
\sigma_{11}^R&=\frac{1}{2\ell}\biggl(i\int_0^1\frac{E\cos\chi+\Omega_S\sin\chi}{E\sin\chi-\Omega_S\cos\chi}d\zeta+\pi\sum_n\frac{2E d}{\chi_n^2}\biggr), \\
\sigma_{12}^R&=\frac{1}{2\ell}\biggl(\int_0^1\frac{\cos(\chi\tilde{x})}{\Omega_S\cos\chi-E\sin\chi}d\zeta+i\pi\sum_n\frac{2E d\cos(\chi_n\tilde{x})}{(-1)^n\chi_n^2}\biggr).
\end{aligned}
\end{equation}
Here, it is understood that after analytic continuation $\Omega_S^2=1-E^2$ and $\chi=2Ed/|\zeta|$. The sums in Eq.~(\ref{self-energycalc}) run over all $n>(2Ed-\chi_0)/\pi$. These are precisely the self-energies that were put in by hand in the numerical procedure described in Sec.~\ref{Methods}. The density of states is read off from the diagonal component of the self-energy,
\begin{equation} \label{DOS}
N=\pi\sum_n\frac{2E d}{\chi_n^2}.
\end{equation}

We now expand the above quantities in the limit $E\to0$, keeping terms to lowest order in both the real and imaginary parts
%DM ?
of the Green's functions.
Because $g$ is symmetric under $\zeta\to-\zeta$ and we are only interested in calculating the LDOS, we are free to take $\zeta>0$. This gives $a_0=1+i(1+2x/\zeta)E$ and $b_0=1+i[1+2(2d-x)/\zeta]E$. With these expansions for $a_0$ and $b_0$, the first-order correction to the Green's function can be expanded further as
\begin{equation} \label{ling}
\delta g=-\frac{1}{2}(\delta a+\delta b)+\frac{i}{2}\biggl[\delta a\biggl(1+\frac{2x}{\zeta}\biggr)+\delta b\biggl(1+\frac{2(2d-x)}{\zeta}\biggr)\biggr]E.
\end{equation}
The correction to the LDOS, $\delta N=\text{Re}\avg{\delta g}$, is therefore
\begin{equation}
\begin{aligned}
\delta N&=-\frac{1}{2}\avg{\text{Re}\,\delta a+\text{Re}\,\delta b}- \\
	&-\frac{E}{2}\avg{\text{Im}\,\delta a\biggl(1+\frac{2x}{\zeta}\biggr)+\text{Im}\,\delta b\biggl(1+\frac{2(2d-x)}{\zeta}\biggr)}.
\end{aligned}
\end{equation}
The self-energies can be expanded as $\sigma_{11}^R=[-iE(1-2d\ln E)+\pi E d]/2\ell$ and $\sigma_{12}^R=1/2\ell$. We note here that all of the above expansions are valid provided that $Ed/\zeta\ll1$. With the above expansions in hand, we continue to calculate the first-order correction to the LDOS.

To simplify the calculation, we assume that the superconductor is perfectly clean. This implies a boundary condition $\delta a=0$ at $x=0$. Given that $\delta a\ll a_0$, the Riccati equation in the normal metal can be linearized,
\begin{equation} \label{lin}
-\zeta\partial_x\delta a+2iE\delta a=2a_0\sigma_{11}^R+\bigl(a_0^2-1\bigr)\sigma_{12}^R.
\end{equation}
Equation (\ref{lin}) admits a solution
\begin{equation} \label{dela}
\delta a=-\frac{\pi E d x}{2\zeta\ell}-\frac{i}{2\zeta\ell}\left(-Ex+2Exd\ln E+\frac{Ex^2}{\zeta}\right).
\end{equation}
Next, we solve a similar linearized Riccati equation for $\delta b$, subject to the specular reflection boundary condition $b=a$ at $x=d$. The solution is given by
\begin{equation} \label{delb}
\begin{aligned}
\delta b&=-\frac{\pi Ed(2d-x)}{2\zeta\ell}
	-\frac{i}{2\zeta\ell}\biggl((2d-x)\times \\
	&\times(-E+Ed\ln E)+\frac{E}{\zeta}(2d-x)^2\biggr).
\end{aligned}
\end{equation}
From Eqs. (\ref{dela}) and (\ref{delb}), we can construct the first-order correction to the Green's function, given in Eq.~(\ref{ling}).

To find $\delta N$, one needs to integrate $\delta g$ over $\zeta$. The real part of $\delta g$ contains a singular term $E^2d^3/\zeta^3\ell$. Because all terms were expanded in the limit $Ed\ll\zeta$, we must introduce a lower cutoff $Ed$ in the angular integral, upon which the corresponding contribution to $\delta N$ becomes independent of $E$:
\begin{equation}
\delta N\sim \int_{E d}^1\frac{E^2d^3}{\zeta^3\ell}d\zeta\sim-d/\ell.
\end{equation}
This term gives the leading correction to the LDOS.

The perturbation theory breaks down when this leading correction becomes of the same order as the LDOS in the absence of impurities, $N\sim Ed$. This linear form is obtained by expanding Eq.~(\ref{DOS}) at small energies and represents the linear suppression of the LDOS produced by the Andreev spectrum. \cite{deGennes:1963} The breakdown of the perturbation theory thus occurs at $E\sim1/\ell$. While this does not prove the existence of a mini-gap, it does show that the system experiences some non-perturbative effect for energies on a scale $E\sim1/\ell$, thus explaining the linear increase in the size of the mini-gap with disorder in the quasi-ballistic limit.

\bibliography{Bib}

\end{document}